\documentclass{article}
\usepackage{amsmath}
\usepackage{fleqn}
\usepackage{xcolor}
\usepackage{graphicx}
\usepackage{tikz}
\usepackage{float}
\usepackage{amsfonts}
\usepackage{amssymb}
\usepackage[toc,page]{appendix}
\usepackage[top=1in,bottom=1in,right=1in,left=1in]{geometry}
\usepackage{amsthm}
\usepackage{subfiles}
\usepackage{lineno}
\DeclareGraphicsExtensions{.pdf,.png}

\usetikzlibrary{positioning}

\usepackage{setspace} 
\usepackage{graphicx}
\usepackage{caption}
\usepackage{subcaption}
\title{Could black hole thermodynamics play a role in black hole mergers?}

\author{George Ruppeiner\footnote{ruppeiner@ncf.edu}\\Division of Natural Sciences\\New College of Florida\\5800 Bay Shore Road\\Sarasota, FL 34243, USA}

\date{\today}

\begin{document}

\maketitle

\begin{abstract}

Gravitational waves detected from binary black hole mergers by the LIGO/Virgo/KAGRA collaboration yield values for both the black hole remnant masses $M$ and spins $a$, with the $169$ spin values collected so far crowding significantly around their average $\bar{a}=0.6869\pm 0.0135$. Could this crowding relate directly to the Davies Point (DP) from black hole thermodynamics? The DP results from the Kerr-Newman black hole model, and has heat capacity diverging at $a=0.68125$, a value very close to the measured $\bar{a}$. In this paper I construct a consistent thermodynamic fluctuation theory for black holes and use it to write the thermodynamic curvature $R$. $R$ immediately yields the correlation length $\xi$. $\xi$ is found to diverge at the DP, and I propose that this divergence brings on critical slowing down that retards the emission of gravitational waves. The spin drifts of the remnants slow in proportion, leading to a piling up of spin value at the DP, as observed. If correct, my work would combine general relativity with black hole thermodynamics in an observational setting.

\end{abstract}

\\

\section{Introduction}

The detection of gravitational waves produced by the merger of binary stellar mass black holes by the LIGO/Virgo/KAGRA collaboration \cite{Abbott2019, Abbott2021, Abbott2023, Abac2025} has opened a new era in the study of black holes. Present methods now allow for the measurement of both the mass $M$ and the spin angular momentum $J$ of the post-merger remnant black holes. Following the lead of Wu {\it et al.} \cite{Wu2022}, I argue that black hole thermodynamics (BHT) \cite{Beckenstein1980, Wald2001} might offer a framework for a striking pattern in the post-merger $(M,J)$ data. Measurements have found that remnant black holes have dimensionless spin values

\begin{equation} a=\frac{|J|}{M^2} \sim 0.7, \end{equation}

\noindent close to the value of the Davies Point (DP) $a=a^*=0.68125$ from BHT \cite{Davies1977}. The DP marks the state where the BHT heat capacity $C_J$ changes sign from negative to positive as $a$ increases it's value through $a^*$, with $|C_J|$ approaching infinity at $a=a^*$.

\par
Is there actually some real causal physical connection between these observed post-merger spin data and the DP? So far as I am aware, this question has not yet been answered definitively. Here I approach this question via the thermodynamic fluctuation theory (TFT).

\par
To determine the $(M,J)$ values of a remnant black hole, the LIGO/Virgo/KAGRA team first measure the gravitational wave ``chirp'' emitted by the source merger event. Measured and theoretically computed chirps are then compared to determine the best match between them. The theoretically computed chirps result from numerical relativity calculations, tabulated in a preprepared catalog. These catalog entries include the computed remnant values $(M,J)$ for each chirp. Numerical relativity consists of numerical solutions of the Einstein field equations, subject to initial conditions. In practice, this powerful solution method works beautifully. However, it offers only numbers, with no general principle for why the $a$ values bunch at the DP. My hope is that bringing in BHT and its phase transitions can lead to some insight.

\par
\begin{figure}[h!]
    \centering
        \includegraphics[width = 0.6\textwidth]{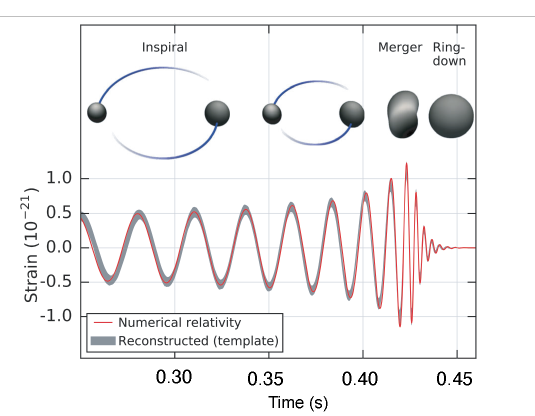}
        \caption{The three stages of the black hole merger process shown for the first detected merger (GW150914). The relevant stage in this paper is the final ring-down stage, where the thermal equilibrium properties of the Kerr model come into play and the construction of the new black hole is completed.}
    \label{Chirp}
\end{figure}

Fig. \ref{Chirp} shows the stages of a black hole merger. The first detected chirp (GW150914) is shown smoothed and superimposed on a matching numerical relativity calculation. This figure, taken from Abbott {\it et al.} \cite{Abbot2016}, shows the dimensionless detector strain versus the time in seconds. Evident are three merger stages, inspiral, merger, and ring-down. The inspiral and merger stages are expected to be highly dynamic. It is only in the final ring-down stage that the dynamics slow and end with the construction of a settled Kerr black hole remnant.

\par
To apply TFT in the black hole scenario, two major challenges must be overcome. The first challenge is to determine which of the three conserved variables $(M,J,Q)$, with $Q$ the charge, actually fluctuate. {\it A priori}, it could be all three, or just one or two, with the other(s) drifting only slowly. A straightforward application of the Kerr-Newman black hole model \cite{Carroll2019, Kerr1963} resolves this issue. I show that it suffices to consider only $\{M,Q\}$ fluctuations. The second challenge is to establish thermal equilibrium between a stellar mass black hole, with it's event horizon at the Hawking temperature $(\sim \mu K)$, and the rest of the universe at the cosmic background temperature ($\sim 3$ K). For this, I employ an intermediate co-rotating system to equalize the temperatures.

\par
The literature of the Davies Point focusses almost exclusively on thermodynamic stability; for a recent approach, see Avramov {\it et al.} \cite{Avramov}. One of my main contributions here is to emphasize that for stability to be physically relevant it should be placed in the larger context of TFT. Stable equilibrium requires not only that the Hessians of the second entropy derivatives be positive definite, but also that the first entropy derivatives (e.g. temperature$\ldots$) must be equal between a system and its environment. Only through the full fluctuation theory approach do essential issues like the proper choice of fluctuating variables and the difficulty of achieving thermal equilibrium come forth for resolution.

\par
The complete remnant formation process must involve as well some dynamical picture. Thermodynamic considerations alone are not sufficient. I propose the idea of a ``critical slowing down,'' taken from the theory of dynamic critical phenomena \cite{Halperin1977, Mann2025} at the DP. An essential element in a critically slowed state is a very large correlation length $\xi$ \cite{Halperin1977}. With large $\xi$, the time required to reach equilibrium over the distance $\xi$ is drastically increased.

\par
Researchers sometimes question the status of the DP as a phase transition, citing a lack of evidence that $\xi\to\infty$ at the DP. One of my contributions here is to bring in the thermodynamic curvature $R$ from the geometry of thermodynamics \cite{Ruppeiner1979} to calculate $\xi$ purely by thermodynamic means. For $\{M, Q\}$ fluctuations, I prove that the resulting $R$ produces an infinite $\xi$ at the DP. Although the DP does not have the clear order parameter expected for a critical point, a diverging $R$ with associate diverging $\xi$ should suffice to produce critical slowing down.

\section{The observational data}

\begin{figure}[h!]
    \centering
    \vspace{-0.75 in}
        \begin{subfigure}[b]{1.0\textwidth}
            \includegraphics[width=\textwidth]{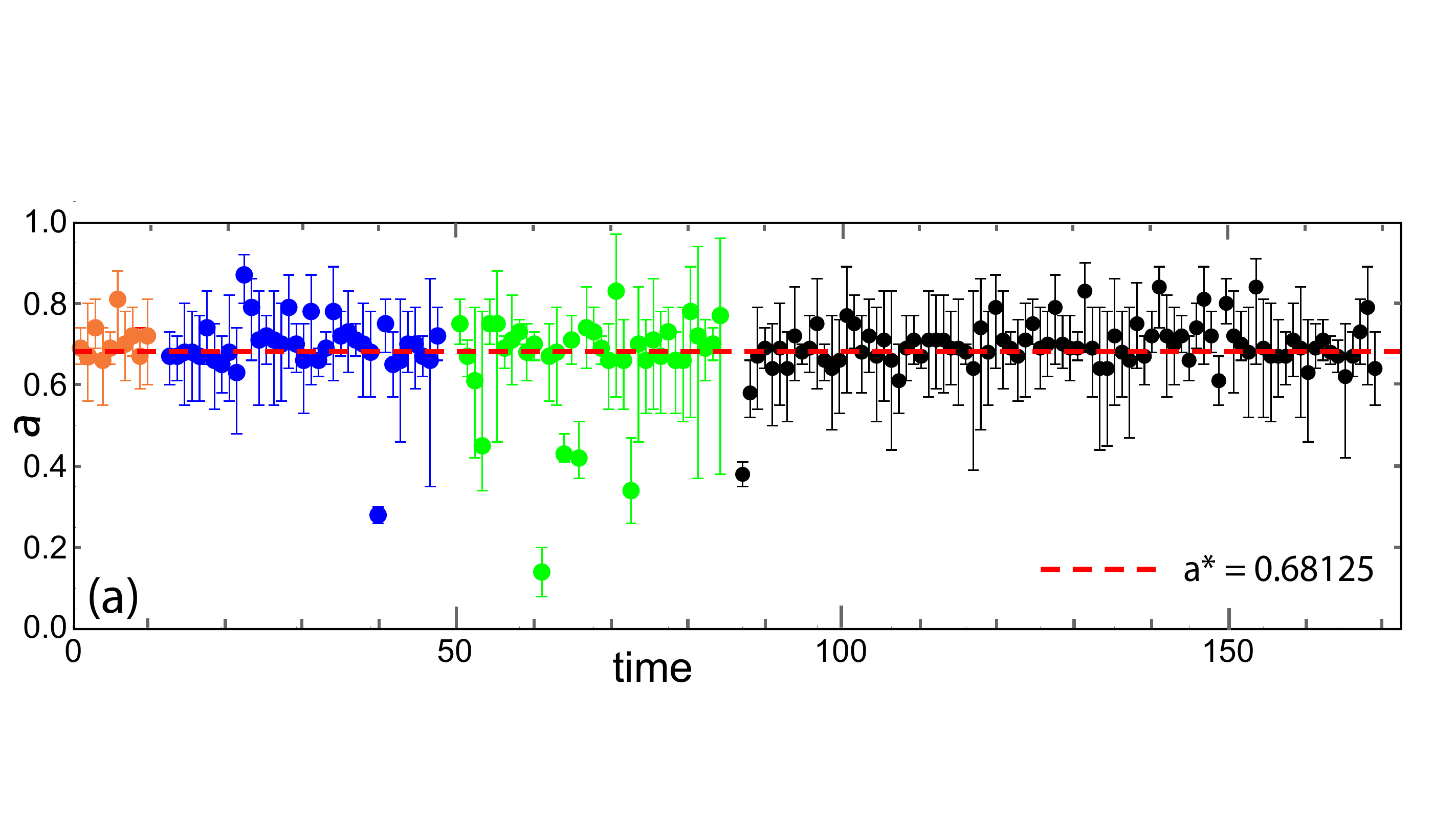}
        \end{subfigure}
    \hfill
    \vspace{-1.5 in}
        \begin{subfigure}[b]{1.0\textwidth}
            \hspace*{-0.21 cm} \includegraphics[width=\textwidth]{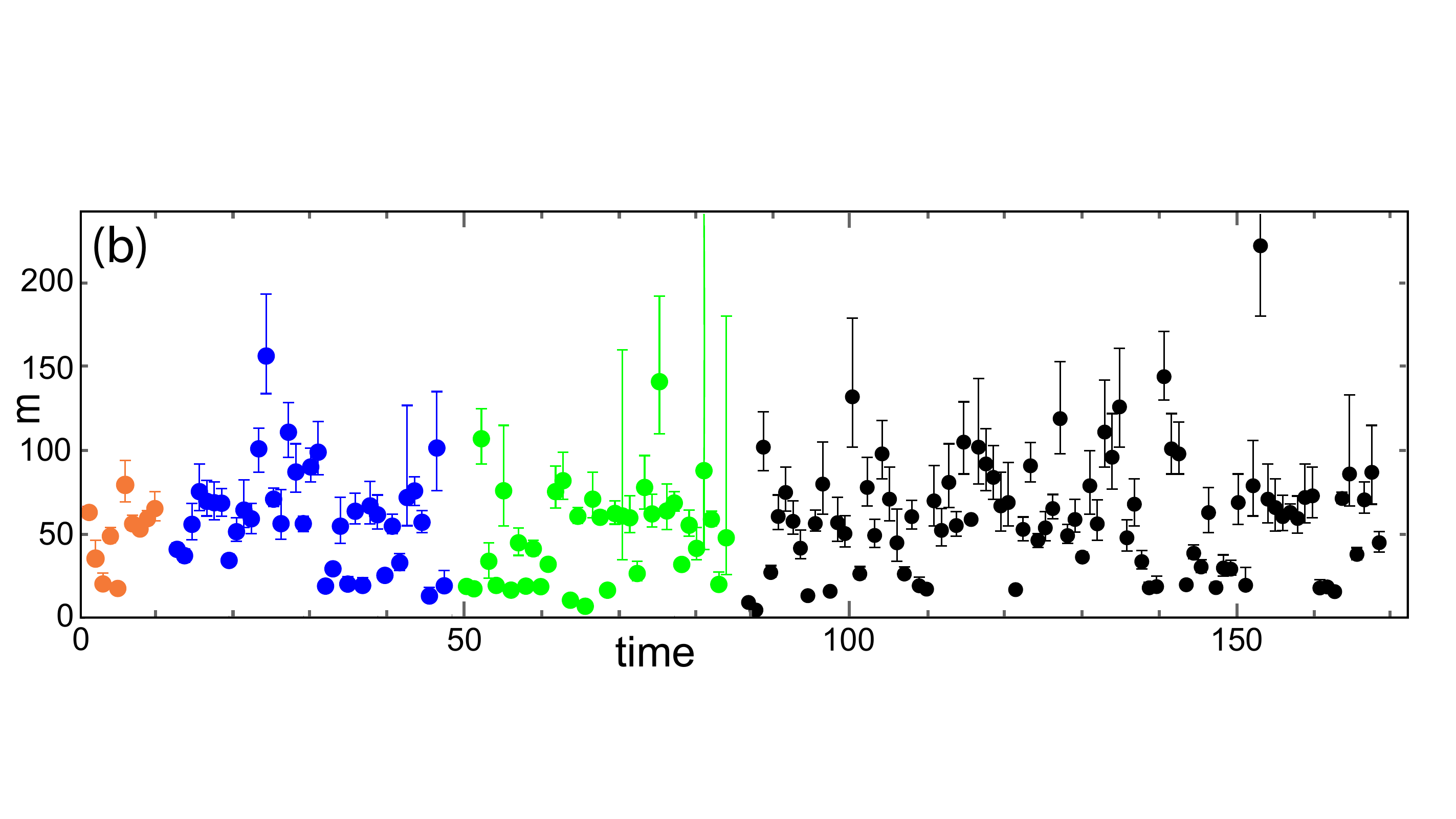}
        \end{subfigure}
     \vspace{-0.8 in}
   \caption{The observed post-merger data: (a) values of the remnant spins $a$ arranged in rough order of increasing time of observation, showing that spin $a$ tends to crowd around the Davies Point at $a=a^*=0.68125$, pictured in dashed red, and (b) values of the remnant masses $m$ (in solar units \(M_\odot\)), which show no particular pattern.}
    \label{DataFigure}
\end{figure}

Look first at the observed remnant spin data motivating this paper. These data were taken in four sets of observing runs \cite{Abbott2019, Abbott2021, Abbott2023, Abac2025}, and are arranged in rough chronological order in Figure \ref{DataFigure}, with points colored brown, blue, green, and black, respectively. The observed post-merger spin data $a$ are dimensionless in the system of units given in \cite{Davies1977}. On physical grounds, $a$ falls in the range $0\le a<1$ \cite{Davies1977}, where $a=0$ corresponds to the non-spinning Schwarzschild black hole, and $a=1$ corresponds to the extremal limit where the black hole spins as fast as is physically possible.

\par
Fig. \ref{DataFigure}(a) shows $a$ for the observed remnant spins reported so far. Visually, there is a clear tendency favoring spin values close to the DP, $a=a^*=0.68125$. The average observed value of $a$ is $\bar{a}=0.6869\pm 0.0135$, where $0.0135$ is the margin of error yielding a 95\% confidence interval. The DP value of $0.68125$ falls into this observed range, so it is consistent with the measured remnant values.

\par
There are $98/169$ data points above the DP, in the BHT stable regime, and $71/169$ data points below the DP, in the BHT unstable regime. $151/169$ of the data points ($90\%$) have their error bars encompassing $a^*$. There are 10/169 points with error bars completely above the DP, and 8/169 with error bars completely below the DP. These observed remnant spin data make a visual case that the DP might have some relevance to the observed data.

\par
For comparison, Figure \ref{DataFigure}(b) shows the corresponding remnant masses $m$, in solar units  \(M_\odot\). The average value of $m$ is $\bar{m}=57 M_\odot$. No particular visual pattern in the distribution of $m$ values is evident.

\par
If BHT requires only that the merger remnants be in the stable thermodynamic regime, then we might expect the data points to skew more above the DP than they actually do. As a measure of this skew, define the skew ratio

\begin{equation} skewratio=\frac{\mbox{number of data points above the DP}}{\mbox{total number of data points}}.\label{794307}\end{equation}

\noindent For the observed remnant spin data, $skewratio=0.58$. If the remnants were always BHT stable, then we might expect future remnant data to tend towards $skewratio\to 1$ as the error bars get smaller. If, on the other hand, the actual clustering of values of $a$ were distributed randomly about the DP then we would expect to find roughly as many values below the DP as above, and $skewratio$ would tend towards $0.5$, close to where they are now.

\par
Ref. \cite{Abbott2019} mentions this crowding of the observed $a$ values, saying ``the medians of all final spin distributions are around approximately $0.7$.'' But they calculate no average of spins. A good source of recent black hole discoveries, including spin values $a\sim0.7$, is the blog by Carlisle \cite{Carlisle2024}.

\section{The relation to thermodynamic stability}

Wu {\it et al.} \cite{Wu2022} brought in the DP, and presented a statistical analysis of the first three sets of observing runs \cite{Abbott2019, Abbott2021, Abbott2023}, concluding that, with high probability, ``thermodynamic stability is robust for $83$ black hole remnants.'' Here, I have added a fourth observing run \cite{Abac2025} (the largest) to the analysis, and did not get quite the skew towards $a>a^*$ that Wu {\it et al.} reported. While the $a$ values in the observed remnant spin data clearly tend to cluster around $a^*$, the points show no overwhelming tendency to pick the stable phase over the unstable one. Nevertheless, thermodynamic stability is essential for BHT, and maybe for real black holes, so I start my discussion of BHT with the theme of stability.

\par
Wu {\it et al.} \cite{Wu2022} employed the Kerr black hole model \cite{Kerr1963} in their analysis. Kerr is a general relativistic, spinning, uncharged, and asymptotically flat model parameterized by $(M, J)$. It is an exact stationary solution to the Einstein field equations. Davies \cite{Davies1977} pointed out that while stellar mass black holes generally have slow spin down rates, the electric discharge rate to $Q=0$ is always very fast. Hence, the charge-less Kerr model seems an excellent first try at modeling real black holes.

\par The Bekenstein-Hawking equation sets the BHT entropy $S=S(M,J)$ to the area $A$ of the event horizon. For Kerr, Davies \cite{Davies1977} wrote:

\begin{equation} S(M,J)=\frac{1}{4}\left(M^2+\sqrt{M^4-J^2}\right).\label{kerrequation}\end{equation}

\noindent If we want to include the charge $Q$, which is necessary below because $Q$ fluctuates, a natural extension of the Kerr model is the Kerr-Newman model. It is an exact stationary solution of the Einstein-Maxwell field equations \cite{Carroll2019} and has \cite{Davies1977}

\begin{equation} S(M,J)=\frac{1}{4}\left(M^2 -\frac{1}{2}Q^2+\sqrt{M^4-J^2-M^2 Q^2}\right).\label{807564757581}\end{equation}

\noindent  I do not use Kerr-Newman in this section, but it does becomes essential when we consider the discussion of which parameters are allowed to fluctuate in TFT.

\par
I use two systems of units in this paper: geometric units with $M$, $J$, $Q$, and $S$ expressed in meters, meters squared, meters, and meters squared, respectively, and standard MKS units for lower case mass $m$, angular momentum $j$, charge $q$ (in esu), entropy $s$, Planck's constant divided by $2\pi$ is $\hbar$, speed of light $c$, Boltzmann's constant $k_B$, and the Planck length $l_p$. The capitalized symbols the gravitational constant $G$ and the solar mass \(M_\odot\) are also in MKS units. We have the conversion factors \cite{Ruppeiner2007}

\begin{equation} m=\left(\frac{c^2}{G}\right) M, \label{7459850}\end{equation}

\begin{equation}  j=\left( \frac{c^3}{G}\right) J, \label{}\end{equation}

\noindent and

\begin{equation}  q=\left(\frac{c^2}{G^{1/2}}\right) Q. \label{}\end{equation}

\noindent To convert the entropy to real units, we have

\begin{equation}\frac{s}{k_B}=S\left(\frac{8\pi}{l_p^2}\right),\label{1608425209}\end{equation}

\noindent where

\begin{equation}l_p=\sqrt{\frac{\hbar G}{c^3}}.\end{equation}

\noindent To be slower than the extremal value of rotation, $J$ must lie in the range $(-M^2,M^2)$. As emphasized by Bekenstein \cite{Bekenstein(1973)}, a theory containing both $\hbar$ and $G$ has a quantum gravity flavor.

\par
The BHT temperature $T$ in the Kerr model is given by

\begin{equation}\frac{1}{T}=\left(\frac{\partial S}{\partial M}\right)_J=\frac{\left(1+\sqrt{1-\frac{J^2}{M^4}}\right)M}{2 \sqrt{1-\frac{J^2}{M^4}}}.\label{rvnvrelp}\end{equation}

\noindent Note that $T\to 0$ as $a\to1$. This marks the extremal limit, where the black hole has its maximum spin rate.

\par
The heat capacity $C_J$ at constant $J$ is given by:

\begin{equation}
	\frac{C_J}{M^2}=\frac{T}{M^2}\left(\frac{\partial S}{\partial T}\right)_J=\frac{1 - a^2 + \sqrt{1 - a^2}}{2 + 2 a^2 -4\sqrt{1 - a^2}},
\label{65890320}\end{equation}

\noindent where I have used the identity

\begin{equation}\left(\frac{\partial S}{\partial T}\right)_J=\left(\frac{\partial S}{\partial M}\right)_J/\left(\frac{\partial T}{\partial M}\right)_J.\end{equation}

\begin{figure}[h!]
    \centering
        \includegraphics[width = 0.6\textwidth]{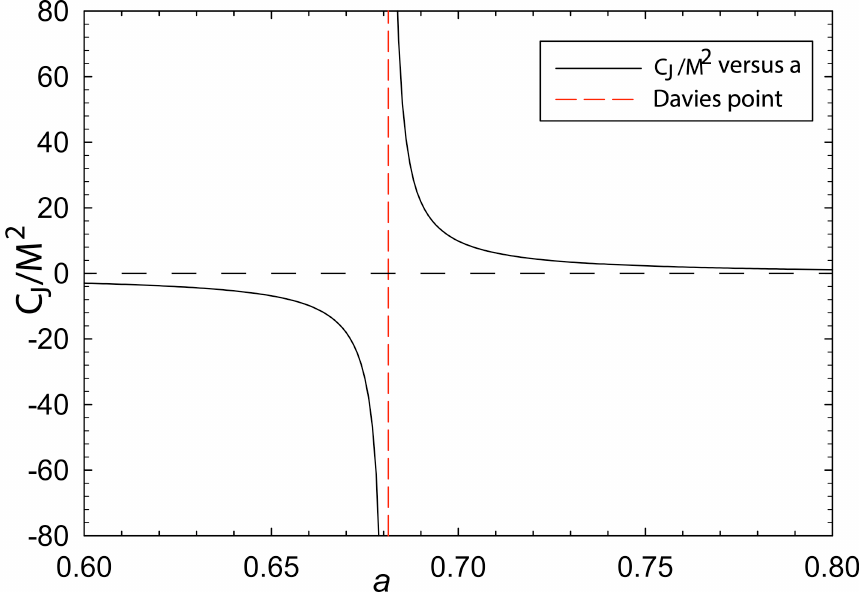}
        \caption{The heat capacity as a function of the spin $a=|J|/M^2$. $C_J$ diverges and changes sign at the Davies Point $a=a^*=0.68125$, indicated with the dashed red line.}
    \label{HeatCapacityGraph}
\end{figure}

\noindent With Eq. (\ref{65890320}) we may show that $|C_J|$ diverges at the DP with $a=a*=\sqrt{-3 + 2 \sqrt{3}}=0.68125$. $C_J$ is negative at all $a<a^*$, and positive at all $a>a^*$; see Figure \ref{HeatCapacityGraph}. Wu {\it et al.} termed these negative and positive sign regimes as thermodynamically unstable and stable, respectively, following textbook terminology \cite{Landau1980,Pathria2011}. However, a positive sign for $C_J$ is only a necessary condition for stability, and it is not a sufficient one. For the full picture, see \cite{Ruppeiner2007}. We will return to this theme in Section 5.

\section{Basic thermodynamic fluctuation theory}

Thermodynamic stability is usually presented in statistical mechanics textbooks \cite{Landau1980, Pathria2011} in the context of thermodynamic fluctuation theory (TFT). TFT requires thermodynamic stability for its application. But, {\it a priori}, this certainly does not force black holes to live only in thermodynamically stable states. One can imagine that a black hole could be held together by its powerful gravity regardless of its BHT stability. The next three sections present the TFT for black hole thermodynamics, a topic with a number of subtleties that have to be considered carefully in physical situations. The ultimate target of this discussion is the thermodynamic curvature $R$, discussed in detail in Section 9. It is via $R$ that TFT enters black hole mergers.

\par
\begin{figure}
\begin{tabular}{c c}
        \includegraphics[width = 0.472\textwidth]{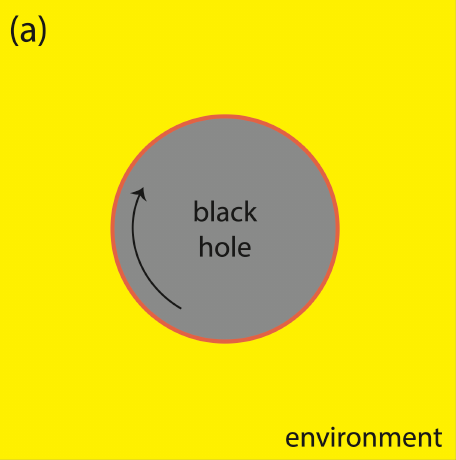}
        \includegraphics[width = 0.466\textwidth]{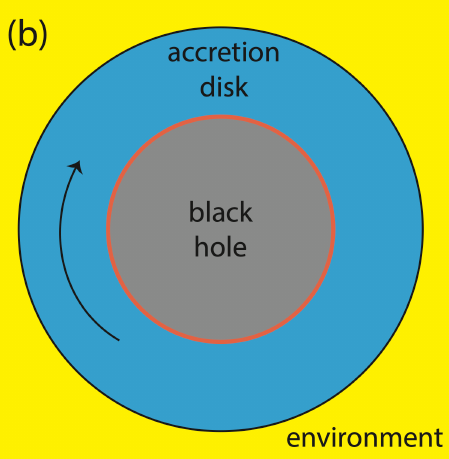}
\end{tabular}
\caption{Sketch of two basic structures framing thermodynamic fluctuation theory for black holes: (a) a spinning black hole surrounded by a non-spinning environment (the rest of the universe), and (b) a spinning black hole connected to the non-spinning environment via an intermediate spinning accretion disc. Fluctuations in the black hole are mediated via Hawking radiation \cite{Hawking1976} at the event horizon, indicated with a red circle. I start my discussion of the Davies Point with scenario (a), and then argue that a correct picture must include the accretion disc in (b) in order to achieve thermal equilibrium at the event horizon.}
\label{6793002234}
\end{figure}

Consider two thermodynamic systems $A$ and $B$. The composite $A$, $B$ system is thermodynamically stable if a steady state equilibrium persists between the two systems, up to small fluctuations about the state with maximum entropy. Pick $A$ as the black hole and $B$ as its environment, which is here the whole rest of the universe. Mediated by Hawking radiation \cite{Hawking1976}, $A$ and $B$ are expected to exchange certain of their conserved parameters. The fluctuating parameters for $A$ are picked from the set $(M,J,Q)$. This basic scenario is shown in Figure \ref{6793002234}(a).

\par
Consider the possibility that one or two of the variables $(M, J, Q)$ fluctuate much more slowly than the other one or two active fluctuating variables, and be effectively fixed in comparison. This leads to seven possible fluctuation cases: fluctuating $\{M, J, Q\}$, $\{J, Q\}$, $\{M, Q\}$, $\{M, J\}$, $\{M\}$, $\{J\}$, and $\{Q\}$. In theory, the detailed dynamics of Hawking radiation could be used to work out which of these scenarios obtains but I am not sure that this has been done. However, if we require that our TFT formulas include a DP, necessary to connect to the observed post-merger spin data in Fig. \ref{DataFigure}(a), then which variables fluctuate clarifies.

\par
But before we resolve this issue, let me explain the basic structure of TFT. For our two system universe $U=A+B$, the system entropies add to give the entropy of the universe:

\begin{equation}S^U(X^A,X^B)=S^A(X^A)+S^B(X^B). \end{equation}

\noindent Here, $X^U$, $X^A$, and $X^B$ are the lists of values of the conserved fluctuating parameters of the systems in play here. The entropies of $U$, $A$, and $B$ are denoted by $S^U(X^A,X^B)$, $S^A(X^A)$, and $S^B(X^B)$, respectively. The conservation laws yield $X^U=X^A+X^B$, which leads to 

\begin{equation} X^B=X^U-X^A, \end{equation}

\noindent with $X^U$ a list of constants. We may then write $S^U$ as a function of just $X^A$:

\begin{equation}S^U(X^A)=S^A(X^A)+S^B(X^U-X^A). \end{equation}

\par
I first take all three parameters $X^A=(M, J, Q)$ to be fluctuating, with straightforward generalization to fixing one or two parameters given later. Einstein's famous TFT equation is \cite{Landau1980, Pathria2011, Einstein1907}

\begin{equation} P(X^A)\propto\exp\left[\left(\frac{8\pi}{l_p^2}\right)S^U(X^A)\right]\label{9691265}\end{equation}

\noindent for the probability density $P(X^A)$ of finding the thermodynamic state of $A$ in some phase space volume element $dM dJ dQ$ located at coordinates $X^A$. The constant in the exponential is the unit conversion factor in Eq. (\ref{1608425209}). This fluctuation formula was written in the black hole scenario by \mbox{\AA}man, Bengtsson, and Pidokrajt \cite{Aman2003}.

\par
Assuming thermodynamic stability, the equilibrium state has maximum total entropy = $S_0$ at coordinate values $X^A=X^A_0$. Expanding $S^U(X^A)$ to second-order about $X^A_0$ in powers of $\Delta X^A = (X^A-X^A_0)$, yields

\begin{equation} S^U(X^A) = S_0 + \frac{\partial S^A}{\partial X^A_i} \Delta X^A_i-\frac{\partial S^B}{\partial X^B_i}\Delta X^A_i + \frac{1}{2}\frac{\partial^2 S^A}{\partial X^A_i \partial X^A_j}\Delta X^A_i\Delta X^A_j+\frac{1}{2}\frac{\partial^2 S^B}{\partial X^B_i \partial X^B_j}\Delta X^A_i\Delta X^A_j + \cdots, \label{6492096}\end{equation}

\noindent where we sum repeated indices $i$ and $j$ over the parameter list $X^A$. The conservation laws yield $\Delta X^B=-\Delta X^A$, and all the derivatives are evaluated at maximum total entropy.

\par
In this expansion, $S_0$ may be absorbed in the normalization constant implicit in Eq. (\ref{9691265}). For the maximum $S^U(X^A)$, the first-order terms in the expansion must add to zero, requiring the equality between the corresponding conjugate parameters of $A$ and $B$:

\begin{equation}\frac{\partial S^A}{\partial X^A_i}=\frac{\partial S^B}{\partial X^B_i}. \label{2657098} \end{equation}

\noindent Generally, we have \cite{Ruppeiner2007}

\begin{equation}\left(\frac{\partial S}{\partial M}\right)_{J,Q}=\frac{1}{T},\,\,\,\left(\frac{\partial S}{\partial J}\right)_{M,Q}=-\frac{\Omega}{T},\,\,\, \mbox{and} \left(\frac{\partial S}{\partial Q}\right)_{M,J}=-\frac{\Phi}{T},\label{278479}\end{equation}

\noindent where $\Omega$ is the black hole angular velocity, and $\Phi$ is the surface electric potential. At the maximum entropy, the second-order terms must add together to be a negative number.

\par
Assume now that the environment $B$ has some Ordinary Thermodynamics (OT) (details unknown and not needed) where both $X^B$ and $S^B(X^B)$ scale up linearly with the volume $V^B$ of $B$. The Hessian of $S^B$ in Eq. (\ref{6492096}) now has the scaling factor $V^B$ one time in it's numerator, and twice in it's denominator. So, as $V_B$ increases, this Hessian goes to zero no matter what the function $S^B(X^B)$ is, while the Hessian of $S^A$ (with $S^A$ given by the Kerr-Newman metric) remains fixed. Hence, with $V^B\gg V^A$, we need only include the known Hessian of $S^A$ in Eq. (\ref{6492096}), and we have, to second order:

\begin{equation} S^U(X) = S_0 + \frac{1}{2}\frac{\partial^2 S}{\partial X_i \partial X_j}\Delta X_i\Delta X_j, \label{7590197}\end{equation}

\noindent where we have dropped the superscript $A$ and have the variables $X_i$ be the fluctuating black hole parameters. A negative definite entropy Hessian matrix is required for thermodynamic stability.

\section{Which conserved parameters actually fluctuate?}

Which conserved parameters in the list $(M,J,Q)$ actually fluctuate? A microscopic dynamical study of Hawking radiation could be tried here, but basic difficulties come up. For example, which types of particles actually fluctuate is not known. There are many possibilities, and we always have the question: are dark matter particles involved? Fortunately, we have an easy way around this. Evaluate all the TFT cases with variables picked from the list $(M,J,Q)$ fluctuating and see which one(s) yield a DP.

\par
But, first, let me add to the picture of the DP that we have used so far, and include the additional rule that the thermodynamic curvature $R$, to be introduced in Section $9$, diverges to infinity at the DP. This sends the correlation length $\xi$ to infinity at the DP, and brings on critical slowing down. The DP has the following five properties: (1) the charge $Q=0$, because black holes shed excess charge quickly \cite{Davies1977, Reynolds2021}, (2) the spin is $a^*$, which is here $0.68125$, (3) the heat capacities $|C_J|$ and/or $|C_Q|\to\infty$, (4) the stability is the full TFT stability, namely, negative definite entropy Hessian for $a>a^*$, and (5) the thermodynamic curvature $R\to\infty$. This last property assures that the DP has critical slowing down, which proves essential to my main argument. We will wait for Sections 9 and 10 for a detailed discussion of $R$. Note, although the average $Q$ is zero at the DP, $Q$ might be fluctuating about this average and we cannot neglect it.

\par
Now search the seven possible candidate cases: $\{M, J, Q\}$, $\{J, Q\}$, $\{M, Q\}$, $\{M, J\}$, $\{M\}$, $\{J\}$, and $\{Q\}$. The cases with a viable DP are suitable for making a connection to the observed remnant spin data, and the cases without one are unsuitable. Ruppeiner \cite{Ruppeiner2007} evaluated all the seven fluctuation cases. Here is a summary of the results: $\{M,J,Q\}$ fluctuations have no regime of full stability, and no DP. They are unsuitable. $\{J,Q\}$ fluctuations are stable everywhere, but have no DP. They are unsuitable. $\{M,Q\}$ fluctuations have a DP and are suitable. $\{M\}$ fluctuations have $R=0$ and are unsuitable. $\{M,J\}$ fluctuations have no regime of stability except for $Q/M$ near $1$. They are unsuitable. In addition, $\{M,J\}$ fluctuations lack a diverging $R$ at $a^*$ \cite{Aman2003}. $\{J\}$ fluctuations are stable everywhere, but have no DP. They are unsuitable. $Q$ fluctuations are stable everywhere, but have no DP. They are unsuitable.

\par
All four cases with fluctuating $J$ are unsuitable. $\{M\}$ meets all of the conditions for a DP other than the curvature condition, so it could be used for calculating Gaussian mass fluctuations as in Section 8. This leaves $\{M,Q\}$ fluctuations as the only fully suitable case. In the $\{M,Q\}$ fluctuations, note that the fluctuations in $Q$ tend to be tiny \cite{Ruppeiner2007}, with a maximum standard deviation $\sqrt{\langle(\Delta Q)^2\rangle}=3.301\,e$ at $J=0$. Here, $e$ is the electron charge.  Despite being featured in a stability discussion in Section 3, $\{M,J\}$ is actually unsuitable because it does not have full thermodynamic stability anywhere near $Q=0$.

\par I add that Sahay \cite{Sahay2017a} gave a general discussion of constrained fluctuations of the type above, with considerations that go deeper than those considered here.

\section{A co-rotating environment is required for thermal equilibrium}

Before returning to the details of the $\{M,Q\}$ fluctuations, I raise a shortcoming with the two system based fluctuation structure in Fig. \ref{6793002234}(a). The Hawking temperature of a roughly solar mass black hole is about $1\mu$K, while the temperature of the whole universe is about the cosmic microwave temperature of $3$ K. It is hard to see how these systems could reach thermal equilibrium with each other, raising a challenge for a TFT based on Fig. \ref{6793002234}(a).

\par
Approaching this thermal equilibrium problem is relatively simple in its basic structure; we introduce an accretion disk located between the black hole and the rest of the outside universe. Let system $A$ be the black hole, let $B$ be an intermediate system between $A$ and $C$, and let $C$ be the non-rotating rest of the universe; see Fig. \ref{6793002234}(b). Rotational drag has us expect $B$ to acquire some measure of co-rotation with $A$. We imagine that the part of $B$ in direct contact with $A$ could now achieve thermal equilibrium with $A$, causing the first order terms in Eq. (\ref{6492096}) to cancel for the $\{M,Q\}$ fluctuation case. $S^A$ is likely to be much more massive than $S^B$ because $A$ will always contain hugely more matter than $B$. This makes the Hessian of $S^A$ in Eq. (\ref{6492096}) much larger than that of $S^B$, and so we could drop the $S^B$ Hessian in Eq. (\ref{6492096}), leaving us with the standard TFT formula Eq. (\ref{7590197}). There seems no need to assume anything about any interaction between $B$ and $C$ as the interface between these two systems would seem to have little to do with $A$ in the short run.

\par
Generally, the theory of accretion disks is difficult. Accretion discs generally become intensely hot from in-falling material. In the centers of distant young galaxies, accretion discs around supermassive black holes are the brightest objects (quasars) in the universe, making thermal equilibrium with the cool event horizon seem unlikely. But a closer look suggests otherwise. Consider the Novikov-Thorne accretion disk model (NT) \cite{Novikov1973}. This model introduces the innermost stable circular particle orbit (ISCO) in the accretion disc. ISCO has radius $r_{\tiny{\mbox{ISCO}}}$, and lies outside the outer event horizon. $r_{\tiny{\mbox{ISCO}}}$ depends on $M$, $a$ and the particle orbital direction, prograde or retrograde. Circular orbits with radii close to, but outside $r_{\tiny{\mbox{ISCO}}}$, have angle $\pi/2$ to the black hole rotational axis. The disc of such orbits is very thin and rapidly rotating.

\par
\begin{figure}[h!]
    \centering
        \includegraphics[width = 0.6\textwidth]{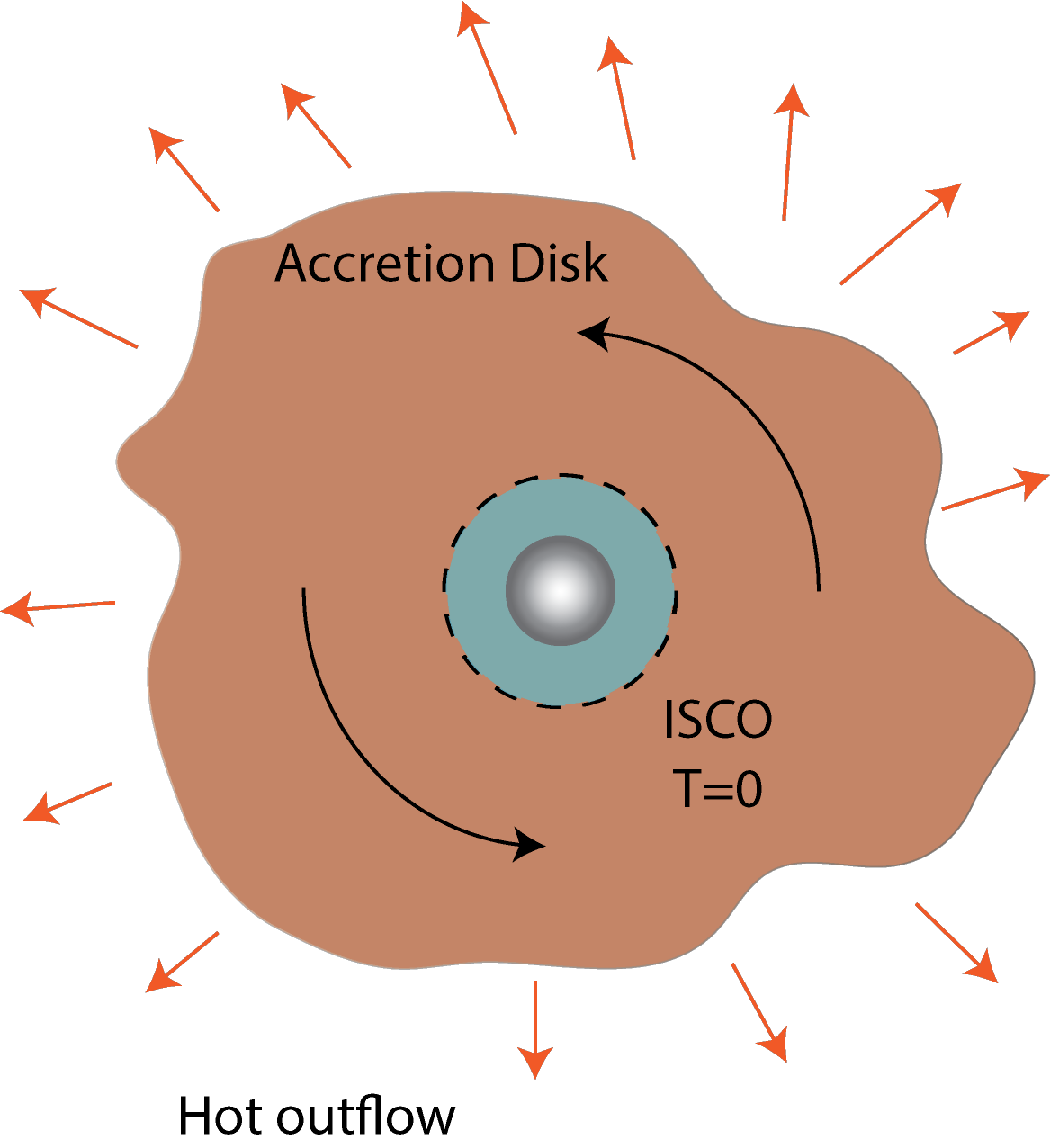}
    \caption{A black hole, its hot accretion disk, and the Inner Stable Circular Orbit (ISCO), where Novikov-Thorne theory predicts zero temperature. This structure allows the black hole's Hawking radiation to equilibrate with the environment inside the ISCO.}
    \label{ISCO}
\end{figure}

NT presents the key finding that as $r \to r_{\tiny{\mbox{ISCO}}}$ from above, the internal stresses, the emitted heat, and the temperature are each predicted to go to zero; see Equation (11) of Reynolds \cite{Reynolds2021}. Inside the ISCO, orbits cease to be stable and are predicted to plunge rapidly to the event horizon because the centrifugal force is no longer strong enough to resist gravity. There seems to be no consensus about the temperature between the ISCO and the event horizon, and this region is not physically observable. Nevertheless, the possibility does arise that the temperature there could be low enough to equilibrate with the Hawking radiation, solving our equilibration problem. See Figure \ref{ISCO} for a diagram of the ISCO.

One might question whether or not there is enough time for the accretion disc to organize itself into the nuanced structure described in this Section in a merger time of about one second. One might also question whether or not the merging remnant would necessarily even have an accretion disk. But let me make an important point: the merger process is far too fast and energetic for any plausible accretion disc to play a direct role! So the character of the accretion disk, NT or otherwise, present or absent, as well as any possible TFT processes taking place outside the black hole have no direct effect on the merger.

Only what is going on in the interior of the developing remnant matters. We need the TFT primarily to develop an indispensable diagnostic tool, the thermodynamic curvature $R$, and this will be done in Section 9. $R$ is employed to determine the correlation length $\xi$ inside the remnant. This is done with the aid of the equation for black hole entropy in Eq. (\ref{807564757581}). $\xi$ is a key measure of dynamical critical slowing down. If $\xi$ is large, the interior dynamics slows, as I argue in Section 10.

\par
The derivation of $R$ could also be done in a purely theoretical context, by picking for the black hole environment a hypothetical infinite universe with temperature set to the black hole Hawking temperature, rotating at the same angular velocity as the black hole, and with electric potential set to zero. This simplified but completely unphysical environment would yield exactly the same results for $R$, Eqs. (\ref{jhryqp}) and (\ref{eoeehedr4}), as the environment used above. The freedom to change environments while leaving the fluctuations the same works because the character of the environment does not directly affect the merger process. In the derivation above, I picked the NT accretion disk because it is physically motivated, and not just some dry mathematical construct.

\section{Have unstable black holes been found?}

Have unstable black holes been discovered? It is possible that they have. Of the observed post-merger spin data shown in Fig. \ref{DataFigure}(a), 8/169 ($\sim 5\%$) of the remnants have both the upper and the lower error bars below the DP, with the lowest spin at ${a}=0.14\pm 0.06$. Of course, this unstable collection is a minority, and it is conceivably spurious.

\par
Another arena to search for unstable, slowly spinning black holes is in pre-merger spin studies \cite{Abbott2019, Abbott2021, Abbott2023, Abac2025}, which employ for analysis a single spin related parameter $\chi_{\mbox{\tiny{eff}}}$, defined in Equation (4) of \cite{Abbott2019}. But $\chi_{\mbox{\tiny{eff}}}$ mixes in the spins of both pre-merger black holes together with their orbital angular momenta and masses. Separating out the values of the individual spins seems difficult.

\par
\begin{figure}[h!]
 \vspace{-0.8 in}
    \centering
        \begin{subfigure}[b]{1.0\textwidth}
            \includegraphics[width=\textwidth]{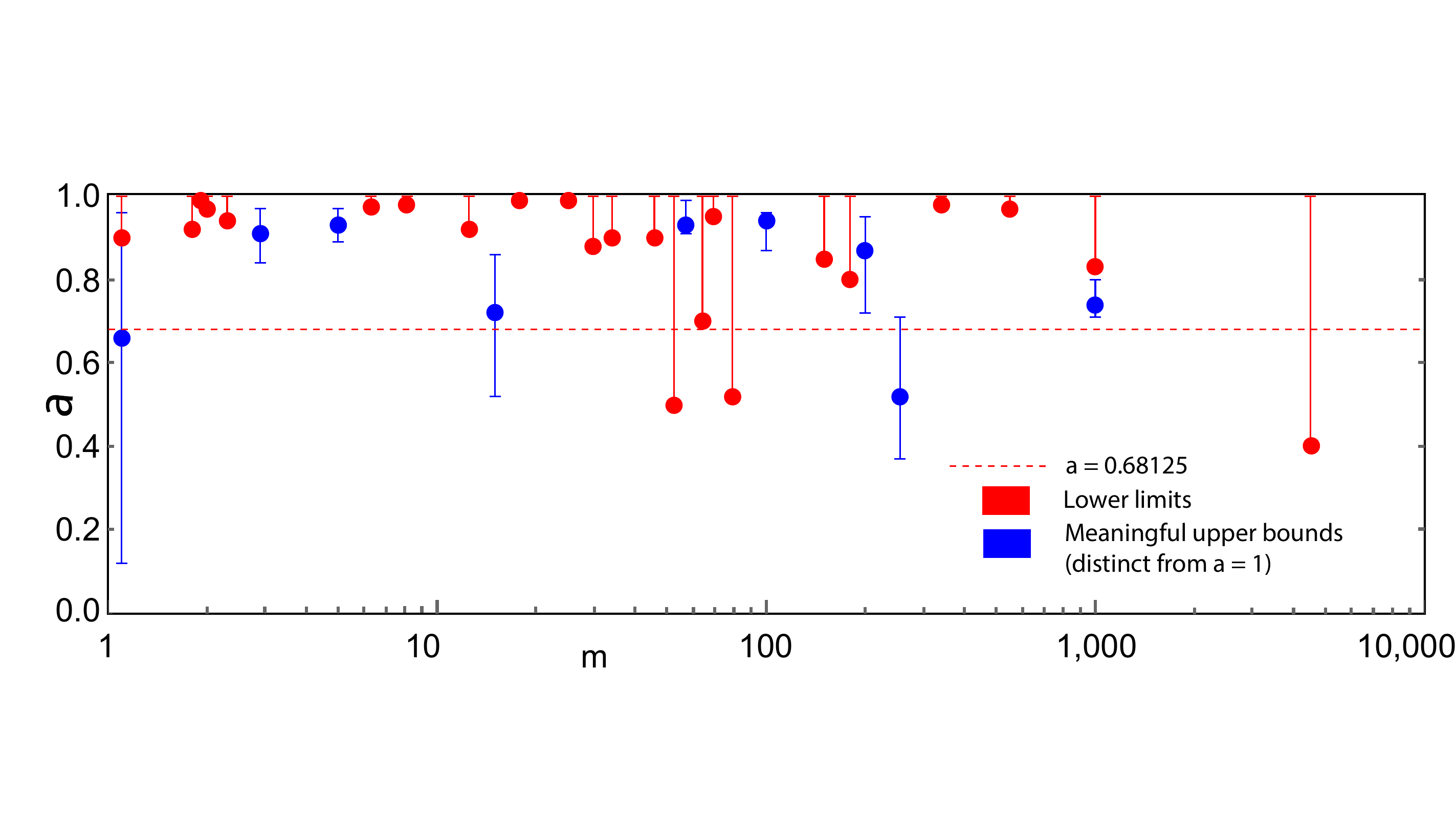}
        \end{subfigure}
     \vspace{-0.8 in}
    \caption{The spin $a$ as a function of the black hole mass $m$ (in units of $10^6$\(M_\odot\)) for supermassive black holes \cite{Reynolds2021}. The blue dots have their upper error bars at $a<1$, and the red dots have their upper errors bars at $a=1$, with their data points placed at their lower error bars.}
    \label{ReynoldsFigure}
\end{figure}

Yet another source of black hole spin measurements is supermassive black holes interacting with their accretion discs. This interaction can be probed with X-ray reflection spectroscopy and with thermal continuum fitting to determine spin. Reynolds \cite{Reynolds2021} reviewed the large literature of measurements of $a$ in these settled supermassive black holes. Their Table $1$, plotted in their Figure 6, shows $32$ spin values for known black hole masses ranging from $1.1\times 10^6 M_\odot$ to $4,500\times 10^6 M_\odot$. I replot these spin results in my Figure \ref{ReynoldsFigure}. The majority of these cases show large spins, many very near the extremal limit $a=1$, and all have their error bars for $a$ either above or encompassing the Davis point.

\par
The supermassive black hole data is in sharp contrast to the observed remnant spin data. A count of the supermassive points shows that their skew ratio $skewratio=27/32=0.84$. If we put all the red points in Fig. \ref{ReynoldsFigure} in the middle of their error bars, rather than at the bottoms, then we get $skewratio=30/32=0.94$. These $skewratio$ values support a stability picture with $skewratio\to 1$, rather than a purely clustering picture with $skewratio\to 1/2$.

\par
We could be of two minds about the supermassive black hole data shown in Fig. \ref{ReynoldsFigure}. We could argue that this is support for a requirement that all BH's must be BHT stable. Alternatively, we could argue that supermassive black holes are naturally spun up very fast by in-falling matter, and that because of their huge masses, they simply spin down very slowly. This would have nothing to do with BHT. Regardless, the supermassive black hole data contrasts sharply with the observed post-merger spin data. This is no surprise because the formation pictures of these respective objects differ substantially, as do their ages, billions of years on the one hand, and roughly a single second on the other.

\section{Black hole thermodynamic mass fluctuations}

\par
Turn now to BHT mass fluctuations, which in the Gaussian approximation diverge at the DP. A stable black hole, with $a>a^*$, spinning down because of various physical drag processes will inevitably encounter the DP at $a=a^*$. Might the black hole disintegrate there from the fluctuation instability? If so, could this potentially help to explain the apparent relative sparsity of black holes in the unstable regime? Below, I write down the BHT Gaussian mass fluctuation formula, but with no observable evidence for any related black hole disintegrations.

\par
Working out fluctuation formulas in MKS units starting from expressions in geometric units requires care. Here are the steps to follow for $\{M\}$ fluctuations: 1) Start with Eq. (\ref{9691265}) in the standard form

\begin{equation} P\propto\exp\left[\frac{\Delta s}{k_B}\right],\label{96915555}\end{equation}

\noindent where $\Delta s$ is the deviation of the entropy (in MKS units) from its maximum value at equilibrium. $\Delta s$ is related to the corresponding $\Delta S$ by the conversion formula Eq. (\ref{1608425209}). Here $\Delta S$ denotes $(S^U(X)-S_0)$ in Eq. (\ref{7590197}). $S$ is given by the Kerr equation Eq. (\ref{kerrequation}). 2) Simple algebra yields the Gaussian expression for the probability distribution for $M$:

\begin{equation} P(M) dM = c \exp\left[-\frac{1}{2}g\,\Delta M^2\right] dM,\label{51053825845}\end{equation}

\noindent where

\begin{equation} g =-\left(\frac{8\pi}{l_p^2}\right)\frac{d^2 S}{d M^2}, \end{equation}

\noindent and $c$ is a normalization constant. 3) Gaussian integration yields $c = \sqrt{g/(2 \pi)}$ and the second fluctuation moment:

\begin{equation} \langle(\Delta M)^2\rangle=\frac{1}{g}.\label{899003758} \end{equation}

\noindent 4) Evaluate $g$ using the Kerr formula Eq. (\ref{kerrequation}) for $S$ and make the replacement $J\to a M^2$. 5) To switch to MKS units, multiply Eq. (\ref{899003758}) twice by the conversion factor in Eq. (\ref{7459850}). For $a>a^*$, this yields \cite{Ruppeiner2007}\footnote{There is a small typo in Table II of ref. \cite{Ruppeiner2007}. In Table II for $\{M\}$ fluctuations with $Q=0$, the term $(1+K)^2$ on the right hand side should be just $(1+K)$, as it is in Table I for $\{M,Q\}$ fluctuations with $Q=0$.}

\begin{equation} \sqrt{\langle(\Delta m)^2\rangle}=\displaystyle 6.1412 \times 10^{-9}\,\mbox{kg}\,\sqrt{\frac{\left(1-a^2\right)^{3/2}}{a^2\left(\sqrt{1-a^2}+3\right)-\sqrt{1- a^2}-1}}. \label{694827013}\end{equation}

\noindent $\{M,Q\}$ fluctuations with $Q=0$ and $a>a^*$ have the same Gaussian $m$ fluctuations as given in Eq. (\ref{694827013}) as well as tiny $q$ fluctuations: $\sqrt{\langle(\Delta q)^2\rangle} \le 3.301 e$. These tiny $q$ fluctuations would seem to have little physical effect, so I will just neglect them and work just with the $\{M\}$ fluctuations. A nice asymptotic approximation for Eq. (\ref{694827013}) is

\begin{equation}\sqrt{\langle(\Delta m)^2\rangle}=\frac{1.6278\times 10^{-9}}{\sqrt{a-a^*}}\mbox{kg}\label{734790678}.\end{equation}

\par
The Gaussian fluctuation theory is limited because the second-order series expansion on which it is based fails if the average $|\Delta m|$ becomes large, as likely happens very near the DP. Big fluctuations require corrections to Gaussian, but naively adding higher-order series terms to the expansion is inadequate since the resulting theory will not be covariant. Ruppeiner \cite{Ruppeiner1995} gave some examples of covariant fluctuation theories in OT. But corrections beyond Gaussian are rarely addressed in the literature, and to my knowledge there has been no discussion of such corrections in the BHT scenario. What happens to mass fluctuations very near the DP is then unknown. In OT, mass fluctuations are generally thought to become very large near critical points. Maybe large mass fluctuations also happen here.

\section{Invariant thermodynamic scalar curvature $R$}

A less problematic measure of fluctuations consists of the invariant thermodynamic Ricci curvature scalar $R$. $R$ is built up from the thermodynamic metric, namely the entropy quadratic form in Eq. (\ref{7590197}) \cite{Aman2003, Ruppeiner1995, Ruppeiner2023}:

\begin{equation}g_{ij}=-\frac{\partial^2 S}{\partial X_i\partial X_j}.\label{gerrefsa}\end{equation}

\noindent The probability of a fluctuation between two thermodynamic states is independent of the thermodynamic coordinate system used for it's calculation, and the quadratic form in Eq. (\ref{7590197}) must transform as a geometric scalar quantity. Hence, $g_{ij}$ above must transform as a second rank metric tensor.

\par
Numerous calculations in statistical mechanic models have confirmed the very effective proportionality,

\begin{equation} |R|\propto\xi^d, \label{mnnfjks}\end{equation}

\noindent where $\xi$ is the correlation length, $d$ is the spatial dimensionality, and the (unwritten) constant of proportionality is of order unity. This program of calculations has also found that $R$ is negative/positive for microscopic interactions attractive/repulsive (with curvature sign convention being negative $R$ for the two-sphere \cite{Ruppeiner1995}). The properties of $R$ are well known, extensively tested, and robust in the OT regime.

\par
$R$ has also seen much attention in BHT. For an early introduction to this topic, see Ruppeiner \cite{Ruppeiner2014}. Many strong results have been obtained by researchers over the years. But, there have been no direct tests of Eq. (\ref{mnnfjks}) for black holes because there are no known statistical mechanical models in play here yet. The basic microscopic structures are simply not yet understood by the scientific community. However, if one believes in the generality of the basic rules of thermodynamics, the same properties should hold in the black hole regime as they do in the OT ones, including Eq. (\ref{mnnfjks}).

\par
The one-parameter BHT cases in this paper: $\{M\}$, $\{J\}$, and $\{Q\}$, have zero intrinsic Riemannian curvatures, and are thus of little physical interest in terms of their thermodynamic curvatures. Likewise, the three parameter BHT case: $\{M,J,Q\}$, has no DP, and no stable fluctuations; see Section 5. So it is also of little physical interest here. This leaves us only with two-parameter geometries, with cases $\{J, Q\}$, $\{M, Q\}$, and $\{M, J\}$. From Section 5, of these, only $\{M,Q\}$ has a DP with a diverging curvature $R$.

\par
$R$ for two-dimensional geometries with coordinates $(x^1,x^2)$, metric elements $(g_{11}, g_{12}, g_{22})$, and $g=g_{11}g_{22}-g_{12}^2$ have curvature \cite{Sokolnikoff}

\begin{equation} \label{RicciScalar}
\begin{array}{lr}
{\displaystyle R= -\frac{1}{\sqrt{g}} \left[ \frac{\partial}{\partial x^1}\left(\frac{g_{12}}{g_{11}\sqrt{g}}\frac{\partial g_{11}}{\partial x^2}-\frac{1}{\sqrt{g}}\frac{\partial g_{22}}{\partial x^1}\right) \right. } \\  \hspace{3.6cm} + {\displaystyle \left. \frac{\partial}{\partial x^2}\left(\frac{2}{\sqrt{g}} \frac{\partial g_{12}}{\partial x^1} -\frac{1}{\sqrt{g}}\frac{\partial g_{11}}{\partial x^2}-\frac{g_{12}}{g_{11}\sqrt{g}}\frac{\partial g_{11}}{\partial x^1}\right)\right].}
\label{errgtrrhtyyjjuy}
\end{array}
\end{equation}

\noindent Evaluating $R$ for $\{M,Q\}$ fluctuations about $Q=0$ with Eq.'s (\ref{807564757581}), (\ref{gerrefsa}), and (\ref{errgtrrhtyyjjuy}) yields the exact:

\begin{equation}R\left(\frac{M^2\pi}{l_p^2}\right)=\frac{\left(\sqrt{1-a^2}-2\right)\left(\sqrt{1-a^2}-1\right)^2}{\left(\sqrt{1-a^2}\right)\left(-a^2+2\sqrt{1-a^2}-1\right)^2}.\label{jhryqp}\end{equation}

\noindent The numerator of this expression has two real roots, both at $a=0$, which are not in the stable zone. The first factor in the denominator has a real root at $a=1$. This root corresponds to the extremal value of $a$, where the temperature $T$ is zero; see Eq. (\ref{rvnvrelp}). This divergence is physically important \cite{Ruppeiner2023}, but it is not in play here. The second factor in the denominator has two positive real roots, both at the DP, $a^*=0.68125$. Hence, $R$ diverges there. A series expansion about $a=a^*$ of Eq. (\ref{jhryqp}) reveals that $R$ diverges at $a^*$ as an asymptotic power law:

\begin{equation} \displaystyle{R\left(\frac{M^2\pi}{l_p^2}\right)=\frac{0.0119661}{(a-a^*)^2}}. \label{eoeehedr4}\end{equation}

\begin{figure}
\begin{tabular}{c c}
        \includegraphics[width = 0.474\textwidth]{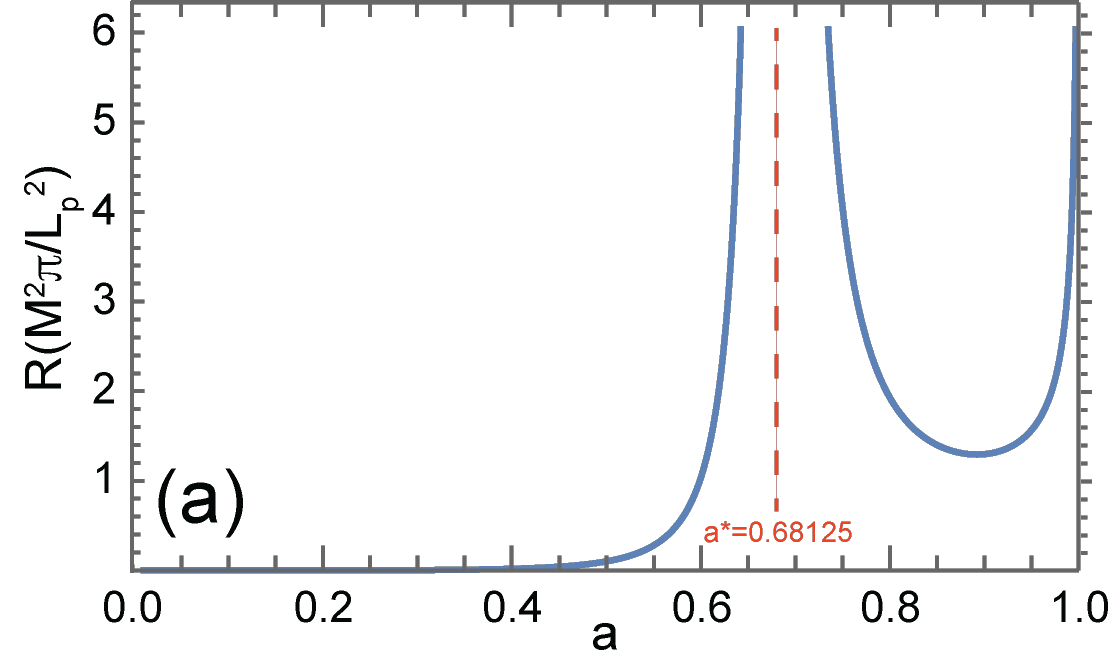}
        \includegraphics[width = 0.466\textwidth]{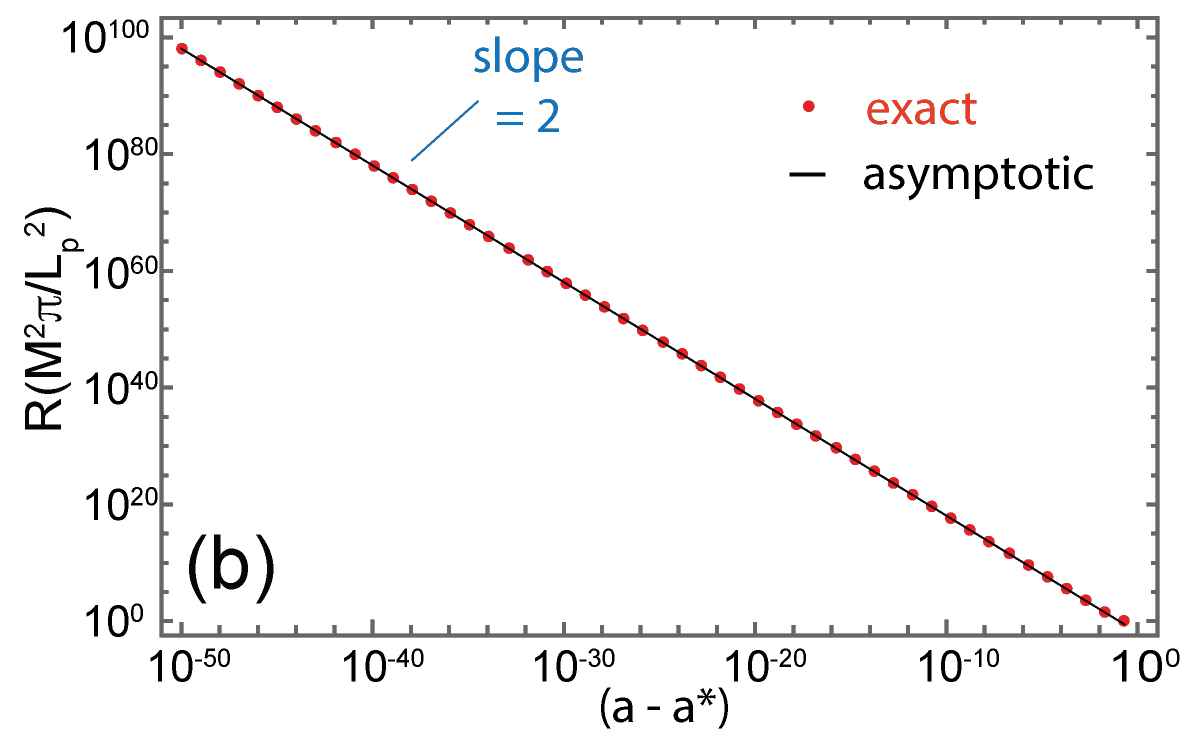}
\end{tabular}
\caption{$R$ for Kerr-Newman $\{M,Q\}$ fluctuations: (a) $R$ over the full range of $a$, and (b) $R$ on a log-log scale demonstrating it's power law divergence at the Davies Point.}
\label{rijenreig}
\end{figure}

\noindent This is a strong power law divergence, with a critical exponent $2$. $R$ is shown in Figure \ref{rijenreig}. $R$ is non-negative, hence displaying the fermionic positive sign characteristic of black holes \cite{Ruppeiner2023}.

For BHT, we expect $|R|\propto\xi^d$. But for lack of consensus microscopic models to calculate with, we cannot determine the scale factor between $R$ and $\xi$. Hence, I have simply left the BHT curvature $R$ dimensionless. This should be adequate for my key argument below. Nevertheless, it would seem that microscopic approaches such as numerical relativity might provide a link between the quantity $R$ computed from BHT and the quantity $\xi^d$ computed from general relativity. This might confirm the power law behavior with exponent $2$ in Eq. (\ref{eoeehedr4}), and yield it's scale factor.

\section{The Davies Point as the remnant attractor state}

Finally, coming back to the main question posed in this manuscript, consider the role of the DP as an attractor state for the remnant spins. Early during the merger, the initial black hole pair has partially combined; see Fig. \ref{Chirp} for the timeline. The absorption of incoming orbital angular momentum has increased the spin of the merging pair and it has a highly distorted, non-axisymmetric shape with bumps and bulges. It is out of conformity with the ``no hair'' theorem of black holes. These deviations in shape from the stationary Kerr black hole get systematically removed during the ring-down, with the emission of enormous energy and angular momentum in the form of gravitational waves (GW).

\par
\begin{figure}[h!]
    \centering
        \includegraphics[width = 0.6\textwidth]{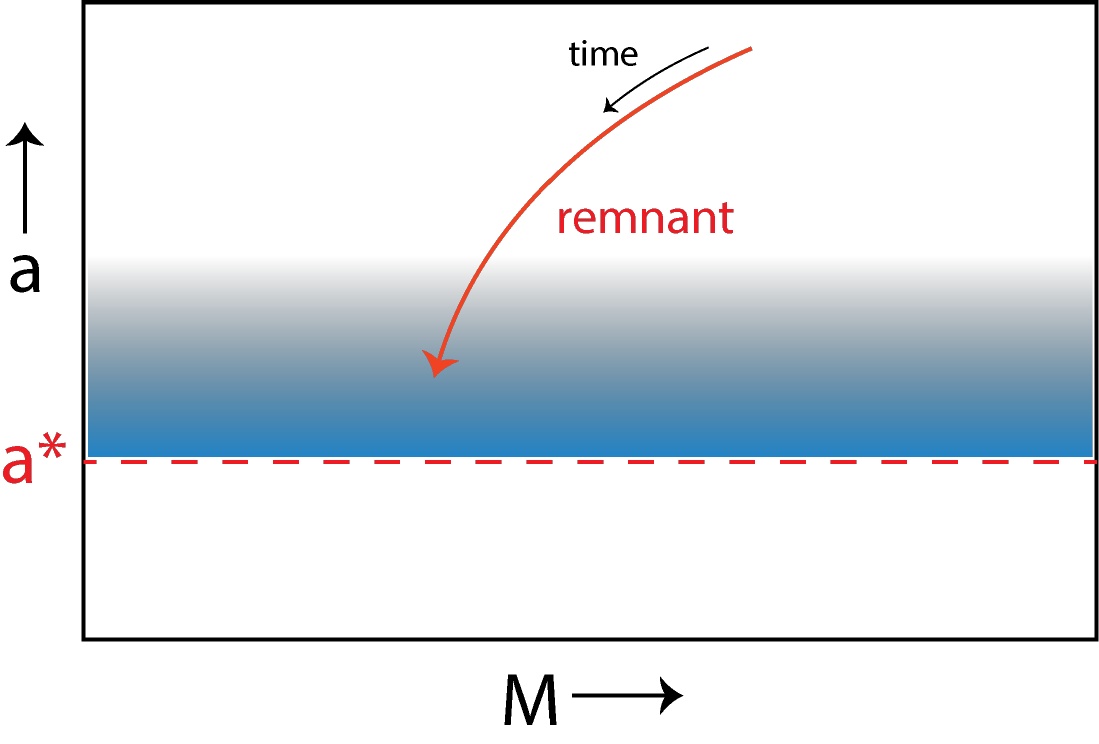}
        \caption{Decreasing remnant values $(M,a)$ as gravitational radiation is emitted on approaching the Davies Point $a^*$. Assuming that the stationary Kerr state is not reached first, emissions slow near $a=a^*$ because of critical slowing down. The blue gradient denotes the equilibrium time for $a>a^*$.}
    \label{SummaryGraph}
\end{figure}

The forming remnant's path in $(M,a)$ space depends both on the initial conditions and on the dynamical process details. A rough sketch of such a path is shown in Figure \ref{SummaryGraph}. With increasing time, this path moves to the left and down with both $M$ and $a$ decreasing. Energy conservation requires decreasing $M$, but one could imagine $a$ increasing or decreasing, or paths starting either above or below the DP at $a^*$. Since most of the remnants in Fig. \ref{DataFigure}(a) end up with spins close to the DP, paths moving towards the DP, either from above or below, seem favored.

\par
In Fig. \ref{SummaryGraph} the pictured descending path moves in the thermodynamically stable region. However, a hypothetical path moving up to the DP would traverse the unstable thermodynamic regime. Is this allowed by the laws of thermodynamics? I take no firm position here, but note that both expressions for $R$ Eqs. (\ref{jhryqp}) and (\ref{eoeehedr4}) are the same regardless of stability. Since the arguments about paths given in this section do not depend in stability, I will not linger on this point. But numerical relativity simulations could offer insight into the nature of these paths.

\par
Let's look at critical slowing down. First, any black hole entropy expression implies a black hole microstructure of entities that collectively determine the BHT variables. Now consider some dynamic process resulting in some small change to $a$. With zero $\xi$ we could change $a$ simply with small adjustments to each individual entity. But with large $\xi$, we must also respect the correlations within each correlated group of entities. This adds a form of resistance to the change in $a$, slowing the process. At the critical point, $\xi$ gets very large and the number of correlated entities to keep organized increases greatly. The dynamic process changing $a$ becomes very slow.

\par
In Fig. \ref{SummaryGraph}, the path descends on the DP from above as the time increases. $R$ from Eq. (\ref{eoeehedr4}) increases as we get closer and $\xi^3$ increases in proportion. The dynamical process slows and comes to a halt at the DP as $\xi\to\infty$. The DP thus acts as an attractor point for the remnant. This explains why the observational data in Fig. \ref{DataFigure}(a) cluster around the DP value $a=a^*$.

\par
It is possible that the remnant achieves the stationary Kerr state before reaching the DP. In this event, the GW emissions cease, with the finished remnant formed early. In Fig. \ref{DataFigure}(a), there seem to be several such orphans both above and below the DP. But these are exceptions, and most of the remnants have piled up near the attractor DP state.

In the picture presented here, the GW emissions from the remnants at the DP likely diminish to the point of undetectability because of critical slowing down. But the stationary Kerr state has probably not been reached. The general question of whether or not a remnant at the end of its chirp has actually reached the stationary Kerr state is a significant research question now being addressed by the LIGO/Virgo/KAGRA collaboration; see Abac {\it et al.} \cite{Abac2026} for a recent reference.

\section{Numerical relativity}

Numerical relativity has seen a role beyond just analyzing the observational gravitational wave data. A large literature exists of purely computer simulations of binary black hole mergers. Since, my focus in this paper is on observational results, I will limit myself to just a few words on computer simulations.

\par
Baumgarte and Shapiro \cite{Shapiro} reviewed the quite involved mathematics of numerical relativity, with attention to merging binary black holes and neutron stars. The simulations start in quasi-equilibrium in the in-spiral phase, enter a chaotic dynamic merger phase, and then quasi-equilibrium in the ring-down phase, before reaching full equilibrium as a stationary Kerr black hole.

\par
Hofmann, {\it et al.}, gave a review of a number of simulation runs, starting with a variety of initial conditions. Most of their reported remnant spin values were at least in the vicinity of $a\sim 0.68$. Perhaps their most famous result reported was for initially equal mass, non-spinning black holes for which they report:

\begin{equation}\bar{a}=0.68646\pm 0.00004, \end{equation}

\noindent very close to the average observed spin value,

\begin{equation}\bar{a}=0.6869\pm 0.0135.\end{equation}

\noindent Hofmann, {\it et al.}, did not speculate on any reason why these $a$ values should cluster so closely together, nor did they raise the close correspondence to the Davies Point value. They just reported their numbers.

\section{Conclusion}

In conclusion, gravitational wave observations have measured the masses $M$ and the spins $a$ of newly formed remnant stellar mass black holes resulting from the mergers of binary black hole systems. I report that the measured spin values cluster around the Davies singularity point (DP), which has $a=a^*= 0.68125$. This suggests a significant connection between the results from gravitational wave measurements and black hole thermodynamics (BHT).

\par
My paper emphasizes this numerical connection to up to the fourth observed remnant spin data set. I have also contrasted the observed remnant spin data with that of the spin data from much older supermassive black holes. The later tend to spin very fast, with their BHT states overwhelmingly in the stable regime above the DP, and no clustering of spin values about any particular number.

\par
Most significantly, I have argued that the DP has its physical roots in thermodynamic fluctuation theory (TFT). It is from TFT that basic concepts such as equilibrium, fluctuations, thermodynamic stability, and thermodynamic curvature come from. In this context, the challenge is to construct a picture with the correct choice of fluctuating variables, as well as thermal equilibrium at the event horizon. I have met this challenge. I have also included the thermodynamic curvature $R$ to establish infinite correlation length $\xi$ and the associated critical slowing down at the DP. I propose that this critical slowing down results in the DP being an attractor state for the binary black hole remnants.

\par
If my method is correct then it brings black hole thermodynamics into the study of binary black hole mergers. This joining of observation, general relativity, and black hole thermodynamics would seem to be a significant advance in the study of black holes.

\section{Acknowledgements}

I thank Andrey Skripnikov for help with the statistics. I also thank the referees for their critical comments on the manuscript. They made my presentation much stronger. I am also very grateful to the late George Skestos for his support, for inspiring me to write about observational gravity, and for asking me many intense questions about gravity.

 \end{document}